\documentclass[pra,aps,twocolumn,superscriptaddress]{revtex4-2}
\usepackage[T1]{fontenc}

\usepackage[utf8]{inputenc}
\usepackage{amsmath,amssymb}
\usepackage{amsfonts}
\usepackage{epsfig,pstricks,graphicx}
\usepackage{bm}
\usepackage{color}
\usepackage{colordvi}

\def\id{{\rm 1\kern-.22em l}}
\newcommand{\expect}[1]{\langle#1\rangle}

\usepackage{enumerate}
\usepackage[capitalize]{cleveref}

\usepackage{float}
\usepackage{lipsum}

\begin{document}

\title{Efficient Characterization of N-Beam Gaussian Fields Through Photon-Number Measurements: Quantum Universal Invariants}
\author{Nazarii Sudak}
\affiliation{Institute of Theoretical Physics, University of Wroclaw, Plac Maxa Borna 9, 50-204 Wrocław, Poland}
\author{Artur Barasi\'nski}
\email{artur.barasinski@uwr.edu.pl}
\affiliation{Institute of Theoretical Physics, University of Wroclaw, Plac Maxa Borna 9, 50-204 Wrocław, Poland}
\author{Jan Pe\v{r}ina Jr.}
\email{jan.perina.jr@upol.cz}
\affiliation{Joint Laboratory of Optics of Palack\'{y} University and Institute of Physics of CAS, Faculty of Science, Palack\'{y} University, 17. listopadu 12, 779 00 Olomouc, Czech Republic}
\author{Anton\' in \v{C}ernoch}
\affiliation{Joint Laboratory of Optics, Institute of Physics of CAS, 17. listopadu 50a, 779 00 Olomouc, Czech Republic}

\begin{abstract}
Quantum universal invariants of general N-beam Gaussian fields are investigated from the point of view of fields' intensity moments. A method that uniquely links these invariants, including the global and marginal fields' purities, to intensity moments is suggested. Determination of these invariants identifies the Gaussian states including their quantum correlations. In particular, the Peres-Horodecki separability criterion is reformulated in terms of quantum universal invariants, and consequently in terms of experimental intensity moments, offering a practical tool for determining the entanglement or separability of these states. The approach is experimentally demonstrated by determining the invariants of noisy symmetric 3-beam Gaussian states using photon-number-resolved measurements. Furthermore, their entanglement properties are analyzed and characterized. 
\end{abstract}

\maketitle

\section{Introduction}
Experimental quantum optics relies on the creation, manipulation, and measurement of the state of light, serving applications from the fundamental scientific research  \cite{WeihsPRL81_1998, ZavattaPRL103_2009, AasiNetPhot_2013} to the development of quantum technologies that exploit the non-classical behavior of light \cite{Bennet_proc, OBrienNatPhot3_2009, TaylorNatPhot7_2013, LiaoPRL120_2018}. Precise characterization of each step in a quantum experiment is crucial for validating theoretical predictions and ensuring proper device operation.
Techniques for quantum-state estimation and quantum-process tomography have been developed for this purpose but, despite significant progress, these methods remain complex and often require precise knowledge of the detector response \cite{CooperNatCom5_2014}. For example, homodyne tomography \cite{Leonhardt1997, Lvovsky2009}, a common technique in quantum optical experiments with continuous-variable states, provides complete characterization of a detected field, though it requires a coherent local oscillator with a varying phase. This introduces serious technical difficulties \cite{Haderka2009}.
Moreover, rapid progress in quantum-enhanced technologies necessitates the use of increasingly complicated systems which demand even more sophisticated measurements.  For example, preparing complex multi-mode entangled states of light distributed among multiple parties is essential for applications in quantum-information processing \cite{DiGuglielmoPRL107_2011} as well as in fundamental physics research. Consequently, the number of measurements, measurement time, and computational effort required for processing tomographic data of multipartite states scale exponentially with the number of parties.

In contrast to homodyne tomography, nowadays we routinely measure photocount histograms in numerous experiments using different types of photon-number-resolving detectors. 
They include, among others, fiber-based photon-number-resolving detectors with time multiplexing \cite{Haderka2004,AvenhausPRL104_2010, SperlingPRL115_2015}, intensified CCD (iCCD) cameras \cite{Haderka2005, PerinaJr2012, PerinaPRApp8_2017} with their spatial multiplexing, hybrid detectors \cite{PerinaAPL104_2014}, silicon multi-pixel detection arrays \cite{RamilliJOSAB27_2010,Chesi2019}, and superconducting bolometers \cite{Harder2016}, each having its own advantages as well as disadvantages.

The relative simplicity of these measurements is a consequence of the fact that the phase of detected light is not captured \cite{Mandel1995}. However, photon-number measurements have been effectively employed to determine many non-classicality witnesses \cite{PerinaJr.2017, PerinaJr.2020, PerinaJr.2022}. Moreover, recent advances have demonstrated that experiments with photon-number-resolving detectors provide both qualitative and quantitative information about the entire hierarchy of bipartite nonclassical correlations in two-beam Gaussian states \cite{BarasinskiPRL2023}. The proposed scheme relies solely on the intensity moments of the detected fields to estimate their global and marginal purities. Although photocount measurements do not allow for the complete reconstruction of the optical fields with their phase properties, substantial information can still be inferred from the values of higher-order intensity moments. This approach has already been successfully validated experimentally in a specific case of two-beam Gaussian fields \cite{BarasinskiPRL2023}. 

Since global and marginal purities are classified as state invariants \cite{Dodonov_2005, SerafiniPRL96_2006} -— quantities that remain unchanged by local unitary transformations —-  the findings presented in \cite{BarasinskiPRL2023} have inspired us for deeper exploration of the general relationship between state invariants and photon-number distributions. More broadly, a fundamental question arises whether a general N-beam Gaussian field can be reconstructed or characterized using solely its photon-number distributions obtained in photon-number measurements. And more specifically, what kind of properties of a quantum state can be inferred from such distributions.

In this paper, we propose a scheme that utilizes the intensity moments of Gaussian optical fields to estimate their quantum universal invariants (QUIs) \cite{Dodonov_2005}. Complete knowledge of these QUIs suffices to determine the symplectic eigenvalues of an optical field, together with its Williamson normal form \cite{WilliamsonAJM58_1936, SimonJMP40_1999}. Our research reveals that some QUIs are exactly determined through photocount measurements. Notably, our scheme uniquely expresses the QUIs involving the global and marginal field purities in terms of higher-order intensity moments. This then  enables direct determination of several entanglement quantifiers, including one- and two-way Gaussian steering \cite{Kogias2015}. The presented approach also allows to estimate the remaining QUIs in the form of their lower and upper bounds. This limitation is a consequence of the absence of phase information. The proposed method is experimentally demonstrated.

The paper is structured as follows. In Sec.~II, $ N $-beam Gaussian fields are described using both their covariance matrix and intensity moments. A general method for the determination of QUIs is presented in Sec.~III. Experimental determination of these QUIs for 3-beam symmetric Gaussian fields starting from the measured photocount histograms is discussed in Sec.~IV. The method is applied for identification of entangled and separable states in Sec.~V. Sec.~VI brings conclusions. The formulas for QUIs of 3-beam Gaussian states are given in Appendix~\ref{AppendixA}. Specific simplified formulas for the residues of some QUIs are contained in Appendix~\ref{AppendixB}. Relations between single- and multimode intensity moments of 3-beam Gaussian fields are given in Appendix~\ref{AppendixC}. Appendix~\ref{AppendixD} brings the model of symmetric 3-beam Gaussian fields.

\section{N-beam Gaussian fields and intensity moments}

We begin with defining the normal characteristic function $C_{\mathcal{N}} (\beta_1,\beta_1^{\ast},\dots, \beta_N, \beta_N^{\ast})$ characterizing a general Gaussian field composed of $ N $ single-mode beams  \cite{Perina1991}
\begin{eqnarray}
&&C_{\mathcal{N}} (\beta_1,\beta_1^{\ast},\dots, \beta_N, \beta_N^{\ast}) = \nonumber\\
&& \exp\Biggl[-\sum_{j=1}^N [B_j|\beta_j|^2+ \left(C_j \beta_j^{\ast 2}/2 + {\rm c.c.} \right) \nonumber\\
&&+ \sum_{j<k}^N \left( D_{jk}\beta_j^{\ast} \beta_k^{\ast}+\bar{D}_{jk}\beta_j\beta_k^{\ast} + {\rm c.c.}\right) \Biggr].
\label{eq:CN}
\end{eqnarray}
In Eq.~(\ref{eq:CN}) the real ($Bj$) and complex ($Cj$, $D_{jk}$, $\bar{D}_{jk}$ ) parameters are defined as
\begin{eqnarray}
    Bj &=& \expect{\Delta \hat{a}_j^{\dagger} \Delta \hat{a}_j}, \quad Cj = \expect{\Delta \hat{a}_j^{2}},\nonumber\\
    D_{jk} &=& \expect{\Delta \hat{a}_j \Delta \hat{a}_k}, \quad 
    \bar{D}_{jk} = - \expect{\Delta \hat{a}_j^{\dagger} \Delta \hat{a}_k},
    \quad j<k
\end{eqnarray}
using the annihilation ($\hat{a}_{j}$) and creation ($\hat{a}_{j}^{\dagger}$) operators of beam $j$, $j =1,2 \dots,N$, and their operator fluctuations $ \Delta\hat{x} \equiv \hat{x} - \langle \hat{x}\rangle $.

The normal characteristic function $C_{\mathcal{N}}$ given in Eq. \eqref{eq:CN} can conveniently be rewritten into the form $C_{\mathcal{N}} (\boldsymbol \beta) = \exp({\boldsymbol \beta}^{\dagger} {\boldsymbol A}_{\mathcal{N}} {\boldsymbol \beta}/2)$ using the covariance matrix ${\boldsymbol A}_{\mathcal{N}}$ related to the normal ordering of the field operators and the column vector $\boldsymbol \beta = (\beta_1,\beta_1^{\ast},\dots, \beta_N, \beta_N^{\ast})^T$. To reveal and quantify the field quantum correlations, we conveniently apply the phase-space approach based on $N$ pairs of canonically conjugated operators $\{\hat{x}_j $ and $ \hat{p}_j\}$, giving the field position and momentum of $ j $th beam, that describes the fields in the symmetric ordering of field operators and corresponds to the Wigner formalism. The reason is technical as we know how to describe and quantify various forms of quantum correlations based on the covariance matrix of a Gaussian state in the symmetric ordering \cite{Adesso2007,Weedbrook2012}. Then, the covariance matrix ${\boldsymbol A}_{\mathcal{S}}$ in the symmetric field-operator ordering is obtained in its block structure as follows:
\begin{eqnarray}
{\boldsymbol A}_{\mathcal{S}} = 
 \begin{pmatrix}
  {\boldsymbol \sigma_{1}} &  {\boldsymbol \varepsilon_{12}} & \dots & {\boldsymbol \varepsilon_{1k}} & \dots & {\boldsymbol \varepsilon_{1N}} \\
  {\boldsymbol \varepsilon_{12}}^T & {\boldsymbol \sigma_{2}} & \dots & {\boldsymbol \varepsilon_{2k}} & \dots & {\boldsymbol \varepsilon_{2N}} \\
  \vdots &      & \ddots & \vdots & & \vdots\\
 {\boldsymbol \varepsilon_{1k}}^T & \dots & 	    &  {\boldsymbol \sigma_{k }} & \dots & {\boldsymbol \varepsilon_{kN}}\\
 \vdots  &  &  & \vdots & \ddots & \vdots\\   		
{\boldsymbol \varepsilon_{1N}}^T & \dots &	& {\boldsymbol \varepsilon_{kN}}^T & \dots &  {\boldsymbol \sigma_{N}} \\   		
 \end{pmatrix},
\end{eqnarray}
where 
\begin{eqnarray}   
 {\boldsymbol \sigma}_j &=& 
 \begin{pmatrix}
  1+2 B_j+2 \Re \{C_j\} & 2 \Im \{C_j\} \\
   2 \Im \{C_j\} & 1+2 B_j-2 \Re \{C_j\}  \end{pmatrix},  \nonumber \\
 {\boldsymbol \varepsilon_{jk}} &=& 
  \begin{pmatrix}
  2 \Re \{D_{jk}-\bar{D}_{jk}\} & 2 \Im \{D_{jk}-\bar{D}_{jk}\} \\
  2 \Im \{D_{jk}+\bar{D}_{jk}\} & -2 \Re \{D_{jk}+\bar{D}_{jk}\} 
   \end{pmatrix}.
\label{8}
\end{eqnarray}
Denoting the symplectic eigenvalues of the covariance matrix ${\boldsymbol A}_{\mathcal{S}}$ as $\nu_j$, we express the Robertson–Schr\"{o}dinger uncertainty relations as $\nu_j \geq 1$, $j=1,\cdots,N$\cite{RobertsonPR34_1929,Schrodinger1930,SerafiniPRL96_2006}. 
The symplectic eigenvalues are themselves symplectic invariants, as they are unchanged by local unitary transformations applied to any of the $ N $ beams. However, they are usually difficult to determine analytically in general for an $N$-beam system. Therefore, a natural choice of these symplectic QUIs is given by the principal minors of the matrix $\Omega {\boldsymbol A}_{\mathcal{S}}$, where the symplectic form $\Omega$ is defined as $\Omega = \bigoplus_{i=1}^{N}
    \begin{pmatrix}
  0 & 1\\
  -1 & 0
\end{pmatrix}$, that is manifestly invariant under symplectic transformations acting by congruence on ${\boldsymbol A}_{\mathcal{S}}$. 
Specifically, following Ref \cite{SerafiniPRL96_2006} and considering a general $n\times n$ matrix $\gamma$, the principal minor $M_k(\gamma)$ of order $k$ of matrix $ \gamma $ is given as the sum of the determinants of all $k \times k$ submatrices of $\gamma$ obtained by simultaneous deleting $n-k$ rows and the corresponding $n-k$ columns. Taking into account the specific structure of the
covariance matrix ${\boldsymbol A}_{\mathcal{S}}$ with $ \hat{x}_j $ and $ \hat{p}_j $ pairs,
the symplectic QUIs $\Delta^N_k$ for $k = 1, \dots, N$ of an $N$-beam state are determined as $\Delta^N_k \equiv M_{2 k}(\Omega {\boldsymbol A}_{\mathcal{S}})$. The quantities $\Delta^N_k$ are also known as QUIs \cite{Dodonov_2005}. The knowledge of the QUIs allows one to determine the symplectic eigenvalues as $N$ solutions of an appropriate system of equations \cite{SerafiniJOSAB24_2007}.

The question is to what extent the symplectic QUIs can be expressed using the (integrated) intensity moments $\expect{W_1^{l_1} \cdots W_N^{l_N}} \equiv \expect{:(a^{\dagger}_1)^{l_1} \cdots (a_N^{\dagger})^{l_N}  a_1^{l_1}   \cdots a_N^{l_N}: }$ obtained from the normal characteristic function $ C_{\mathcal{N}} $ along the formula
\begin{eqnarray}
  &\expect{W_1^{l_1} \cdots W_N^{l_N}} =  & \nonumber\\    
  &  \left. \frac{\partial^{(2\sum_j l_j)} C_{\mathcal{N}} (\beta_1,\beta_1^\ast,\dots, \beta_N, \beta_N^\ast) }{ \partial^{l_1} \beta_1 \partial^{l_1} (-\beta_1^*) \cdots \partial^{l_N} \beta_N \partial^{l_N} (-\beta_N^*)} \right |_{\beta_1 = \dots = \beta_N^\ast = 0}.
\label{eq:intensity}
\end{eqnarray}
We note that the intensity moments are in fact the normally ordered photon-number moments and they are obtained from the measured photon-number moments via the Stirling numbers \cite{Gradshtein2000}.  

Considering the intensity moments up to second order, we reveal the following relations
\begin{eqnarray}     
 \expect{W_j} &=& B_j, \nonumber \\
 \expect{W_j^2}  &=& 2 B_j^2+|C_j|^2, \nonumber \\
 \expect{W_jW_k} &=& B_j B_k + |D_{jk}|^2+|\bar{D}_{jk}|^2.
\label{eq:intensity2}
\end{eqnarray}
where $j,k =1,2 \dots,N$ and $j<k$. 
The intensity moments are defined as ($\sum_j l_j$)th-order cumulative powers of intensities describing the individual beams [as seen in their definition in Eq.~\eqref{eq:intensity}]. They depend on the parameters of interest. However, they do not provide complete information about these parameters, since the photocount measurements are phase-insensitive.

\section{Method for determining quantum universal invariants}
\label{Method}

Before discussing a general scheme, let us first consider a simple but important example. For this purpose, let us take the symplectic QUI of a 1-beam state described by ${\boldsymbol A}_{\mathcal{S}}$
\begin{equation}   
    \Delta^1_1 = \det \sigma_1 = 1 + 4 B_1 + 4 B_1^2 - 4 |C_1|^2.
\label{7}
\end{equation}
As one can see, the QUI $\Delta^1_1$ is represented as a second-order polynomial in the Gaussian-state parameters. This is a general observation resulting from the definition of $\{\Delta^N_k\}$ where the determinants of the $2 k \times 2 k$ submatrices of ${\boldsymbol A}_{\mathcal{S}}$ contain the products of $2 k$ non-zero elements of ${\boldsymbol A}_{\mathcal{S}}$ and, according to Eq.~(\ref{8}),
the products of the Gaussian-state parameters up to the  $2 k$-th order. To express $\{\Delta^N_k\}$ in terms of intensity moments, one has to consider in general a linear combination of all intensity moments up to the $2 k$-th power. Thus, the general form of the QUI $ \Delta^1_1 $ is expressed as the following linear combination:
\begin{eqnarray}
\Xi_1&=& e_0 + e_1 \expect{W_1} + e_2 \expect{W_1}^2 + e_3 \expect{W_1^2},
\end{eqnarray}
where the looked-for coefficients are denoted as $e_i$, $i=0,1,2,3$. 
If there exists a suitable linear combination $\Xi_1$ such that $\Xi_1
= \Delta^1_1$, the coefficient $ \delta $,
\begin{eqnarray}
\delta &\equiv& \Delta^1_1-\Xi_1 = 1 - e_0 + (4 - e_1) B_1  \nonumber\\
&+& (4 - e_2 - 2 e_3) B_1^2 - (4 + e_3) |C_1|^2, 
\end{eqnarray}
has to equal $0$ independently of the value of the Gaussian-state parameters $B_1$ and $C_1$. 
This leaves us the following set of equations:
\begin{eqnarray}   
1 - e_0 &=& 0, \nonumber\\
4 - e_1&=& 0,\nonumber\\
4 - e_2 - 2 e_3&=& 0,\nonumber\\
4 + e_3&=& 0.
\label{10}
\end{eqnarray}
The solution of Eqs.~(\ref{10}) is easily found as $e_0 = 1$, $e_1 = 4$, $e_2 = 12$, $e_3 = - 4$. It
provides the formula~(\ref{7}) for the symplectic QUI $\Delta^1_1$ in the form:
\begin{equation}
    \Delta^1_1 = 1 -4 \expect{W_1} + 12 \expect{W_1}^2 -4 \expect{W_1^2}.
\label{eq:delta11}
\end{equation}

This example, when generalized to an arbitrary symplectic QUI $\Delta^N_k$, gives us the general approach defined as follows.
\begin{enumerate}
    \item Determine a symplectic QUI $\Delta^N_k$ of an $N$-mode Gaussian state in terms of its parameters.
    \item Write down a formula $\Xi_k$ for a general linear combination of all intensity moments up to the $2k$-th order.
    \item Using Eq. \eqref{eq:intensity}, derive the set of linear algebraic equations like that in Eq.~(\ref{10}) whose solution guarantees $\delta \equiv \Delta^N_k-\Xi_k = 0 $.
    \item If such solution exists, the  $\Delta^N_k$ is uniquely determined in terms of intensity moments.
\end{enumerate}

\begin{table}    
\begin{ruledtabular}
		\begin{tabular}{ccc} 
		Sym. Inv. &  Lin. Comb. & Residue \\ \hline
		$\Delta^1_1$   & Eq. \eqref{eq:delta11} & -- \\
		$\Delta^2_1$   & Eq. \eqref{eq:delta21} & $8 (|D_{12}|^2-|\bar{D}_{12}|^2)$ \\
		$\Delta^2_2$   & Ref. \cite{BarasinskiPRL2023} & -- \\
		$\Delta^3_1$   & Eq. \eqref{eq:delta31} & $8 \big(|D_{12}|^2-|\bar{D}_{12}|^2 + |D_{13}|^2-|\bar{D}_{13}|^2$ \\
                        & & $+ D_{23}|^2-|\bar{D}_{23}|^2\big)$ \\
		$\Delta^3_2$   & Eq. \eqref{eq:delta32} & Eq. \eqref{eq:R32} \\
  		$\Delta^3_3$   & Eq. \eqref{eq:delta33} & -- \\
		\end{tabular}
	\end{ruledtabular}	
\caption{\label{tab1} Reference table for the formulas for QUIs $ \Delta^N_k $, $ k=1,\ldots, N $ of $ N $-beam Gaussian states for $ N=1,2,3 $ including their residues.}
\end{table}

Applying this procedure, one can address the symplectic QUIs of general Gaussian fields. The symplectic QUI $\Delta^2_2$ of a 2-beam Gaussian field has already been discussed in Ref.~\cite{BarasinskiPRL2023}. Similarly, in Tab.~\ref{tab1} we summarize the results for the Gaussian fields composed of up to 3 beams. 

This procedure also reveals that a given symplectic QUI cannot be expressed as a suitable linear combination of intensity moments. The QUI $\Delta^2_1 = \det \sigma_1 + \det \sigma_2 + 2 \det \varepsilon_{12}$ represents the simplest example of this kind, where the set of equations like (\ref{10}) does not have a solution. There occur contradictory equations in this set. Ignoring them, we succeed to partially express the QUI $\Delta^2_1  $ in terms of intensity moments:  
\begin{eqnarray}     
    \Delta^2_1 &=& 2 + 4 \expect{W_1} + 12 \expect{W_1}^2 - 4 \expect{W_1^2} + 4 \expect{W_2} \nonumber\\ 
    &+& 12 \expect{W_2}^2 - 4 \expect{W_2^2} - \Delta^2_{1,r}.
    \label{eq:delta21}
\end{eqnarray}
In Eq.~(\ref{eq:delta21}), $\Delta^2_{1,r} = 8 (|D_{12}|^2-|\bar{D}_{12}|^2)$ stands for the residue which cannot be expressed solely via the intensity moments. Considering Eq.~\eqref{eq:intensity2} one can only estimate its upper bound using the following inequality:
\begin{eqnarray}    
\Big||D_{12}|^2-|\bar{D}_{12}|^2 \Big| \leq |D_{12}|^2+|\bar{D}_{12}|^2 \nonumber\\
= \expect{W_1W_2}-\expect{W_1} \expect{W_2}.
\label{eq:res1}
\end{eqnarray}

To summarize, our method either reveals the symplectic QUIs in terms of intensity moments or identifies and quantifies the irreducible remainder/residue, as documented in Tab.~\ref{tab1}. When there exists a residue, its lower and/or upper bounds can be set using the intensity moments. Alternatively, we can establish these bounds employing various relations among the QUIs, as discussed in Ref. \cite{SerafiniPRL96_2006}. For example, using the relation $\Delta^2_2-\Delta^2_1+1\geq 0$ one can find $\Delta^2_{1,r} \geq -\Delta^2_2 + \Delta^2_{1,w}-1$, where we introduce $ \Delta^2_{1,w} = \Delta^2_1 + \Delta^2_{1,r}$ as a part of $ \Delta^2_1 $ expressible through intensity moments  [see Eq.~\eqref{eq:delta21}]. Consequently, one has 
\begin{eqnarray}    
 &\max\{-\Delta^2_2 + \Delta^2_{1,w}-1 ,  -8\expect{W_1W_2}+8\expect{W_1} \expect{W_2}  \}   \leq & \nonumber \\
 & \Delta^2_{1,r} \leq 8\expect{W_1W_2}-8\expect{W_1} \expect{W_2}.& 
\label{eq:boundaries}
\end{eqnarray}
The analysis of the residue of a QUI can also be simplified by imposing specific properties to the analyzed state. For example, $|\bar{D}_{jk}|^2=0$ holds for a general noisy twin beam \cite{PerinaJr.2019}.

The results of a systematic approach to derive the formulas for the QUIs of 1-, 2-, and 3-beam Gaussian states including the accompanying residues are summarized in Tab.~\ref{tab1}. These formulas are assumed to be the most useful in practical applications. Detailed calculations are given in Appendices \ref{AppendixA} and \ref{AppendixB}, together with the estimations of the lower and upper bounds of the derived residues. To demonstrate the usefulness and strength of the derived formulas, we apply them to the experimental 3-beam symmetric Gaussian states.

\begin{figure*}  
\raggedright
  \begin{center}
   \input{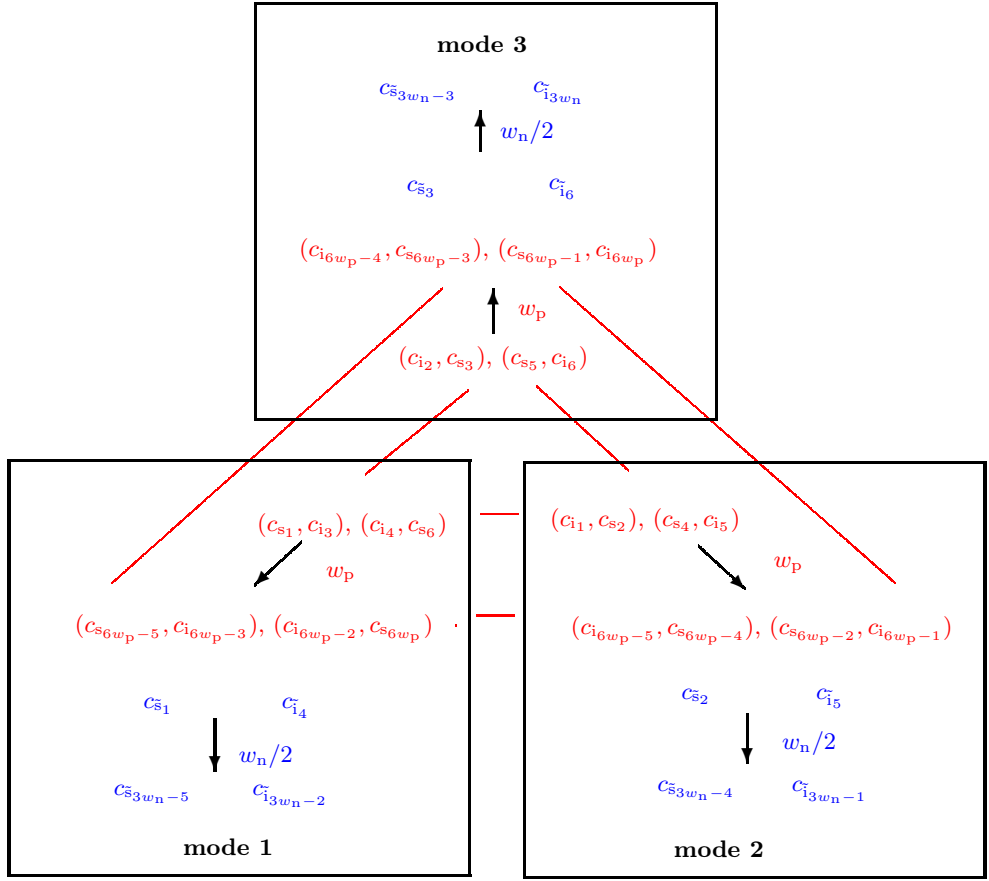}
  \end{center}
 \caption{Scheme for composing symmetric 3-beam Gaussian states from a sequence of weak twin beams
  detected by 2 single-photon sensitive APDs providing the photocounts in the signal and
  idler detection channels. For details, see the text.}
\label{fig1}
\end{figure*}

\section{Experimental determination of quantum universal invariants for three-beam Gaussian states}
\label{Experiment}

The above derived formulas for the QUIs are applied to the experimental symmetric 3-beam Gaussian states.
The analyzed symmetric  3-beam  Gaussian states were prepared using the method of compound beams \cite{PerinaJr.2021}. In this method, the fields are constructed from the experimental data representing the photocounts of the signal and idler beams, obtained by detecting a long sequence of identical weak twin beams (TWBs) originating in pulsed spontaneous parametric down-conversion. This detection was performed using two single-photon-sensitive avalanche photodiodes (APDs). 
The APDs were illuminated by the signal and the idler beams containing on average around 0.1 photons per pulse (detection window), i.e. under the condition where the pile-up effect in the detection by an APD is relatively weak. Moreover, the distortion of the photocount histogram caused by this effect is involved in the detector detection matrix [see Eq.~(\ref{16}) below] and so it is (partially) corrected when reconstructing the photon-number distribution \cite{PerinaJr.2021}.  
The obtained photocount data form two synchronized channels (one for the signal photocounts and the other for the idler photocounts) with $ N_M = 695 \times 10^6 $ measurement outcomes. Each realization of weak TWB results in one of four possible results: no detection at either APD, detection only at the signal APD, detection only at the idler APD, or coincidence detection at both APDs. More details about the experiment can be found in \cite{BarasinskiPRL2023}.

The construction of the analyzed states and preparation for their analysis are composed of the following four consecutive steps:
\begin{itemize}
 \item Grouping and symmetrizing the photocount measurements characterizing the weak twin beams according to the structure of the investigated fields.
 \item Maximum-likelihood reconstruction of the obtained photocount histograms to arrive at the corresponding photon-number distributions using the known detector parameters.
 \item Photon-number moments of the reconstructed distributions are determined and transformed into the normally-ordered photon-number, i.e. intensity, moments.
 \item Choosing a suitable number of independent modes involved in the reconstructed field, the field intensity moments are reduced to those characterizing one typical independent mode.  
\end{itemize}    

In detail, the scheme for building symmetric 3-beam  Gaussian states from two channels of the signal and idler photocounts denoted as $ c_{{\rm s}_j} $ and $ c_{{\rm i}_j} $, $ j=1,2,\ldots $, is drawn in Fig.~\ref{fig1}.
These states consist of the correlated and the noise components. The basic building block of the correlated component is made up of six double (signal and idler) windows, each containing $ (c_{{\rm s}_j}, c_{{\rm i}_j}) $ photocounts. These photocounts are assigned to modes 1, 2, and 3 according to the scheme shown in Fig.~\ref{fig1}. Due to asymmetry caused by different detection efficiencies and dark-count rates of the APDs in the signal and the idler detection windows, symmetric two-beam correlations are formed by pairing double windows where the roles of the signal and idler photocounts are alternated. For example, when building symmetric correlations between modes 1 and 2, we use double windows $ (c_{{\rm s}_1}, c_{{\rm i}_1}) $ and $ (c_{{\rm s}_4}, c_{{\rm i}_4}) $ [see the scheme in Fig.~\ref{fig1}]. This construction ensures symmetry in the basic building blocks of the correlated component and, consequently, symmetry in the correlated part of the analyzed state, which in general contains $ w_{\rm p} $ basic correlated building blocks.

The 3-beam Gaussian states also include the noise components, with photocount numbers derived from the signal and idler channels using $ w_{\rm n}/2 $ signal and $ w_{\rm n}/2 $ idler detection windows for each beam
($ w_{\rm n} $ is assumed to be even). This sequence of photocounts, for which we reserve $3w_{\rm n} $ 
detection windows in both the signal and idler channels (only one half of them is in fact used), is taken from a different part of the (synchronized) signal and idler channels compared to the sequence of length $ 6w_{\rm p} $ used for the correlated component. A given realization of the state, starting from position $ j $ of the double window for the correlated part and $ \tilde{j} $ for the noise part, is characterized by $ c_k $ photocounts in beam $ k $ for $ k=1,2,3 $ determined as:
\begin{eqnarray}  
 c_{1}(j,\tilde{j}) &=& \sum_{k=1}^{w_{\rm p}} [c_{{\rm s}_{j+6k-6}} + c_{{\rm i}_{j+6k-4}}
  + c_{{\rm i}_{j+6k-3}} + c_{{\rm s}_{j+6k-1}}] \nonumber \\
  & &  + \sum_{k=1}^{w_{\rm n}/2} [ c_{{\rm s}_{\tilde{j}+6k-6}} + c_{{\rm i}_{\tilde{j}+6k-3}} ], \nonumber \\
 c_{2}(j,\tilde{j}) &=& \sum_{k=1}^{w_{\rm p}} [c_{{\rm i}_{j+6k-6}} + c_{{\rm s}_{j+6k-5}}
  + c_{{\rm s}_{j+6k-3}} + c_{{\rm i}_{j+6k-2}}]  \nonumber \\
  & & + \sum_{k=1}^{w_{\rm n}/2} [c_{{\rm s}_{\tilde{j}+6k-5}} + c_{{\rm i}_{\tilde{j}+6k-2}} ],
  \nonumber \\
 c_{3}(j,\tilde{j}) &=& \sum_{k=1}^{w_{\rm p}} [c_{{\rm i}_{j+6k-5}} + c_{{\rm s}_{j+6k-4}}
  + c_{{\rm s}_{j+6k-2}} + c_{{\rm i}_{j+6k-1}}]  \nonumber \\
  & &   + \sum_{k=1}^{w_{\rm n}/2} [ c_{{\rm s}_{\tilde{j}+6k-4}} + c_{{\rm i}_{\tilde{j}+6k-1}} ].
\label{15}
\end{eqnarray}
By incrementing $ j $ step by step and selecting $\tilde{j} $ from different regions of the photocount channels, we can generate nearly $ N_M $ realizations of a symmetric 3-beam Gaussian state characterized by the numbers $ w_{\rm p} $ and $w_{\rm n} $. These realizations contribute to a photocount histogram $ f(c_1,c_2,c_3;w_{\rm p},w_{\rm n}) $ which gives the probability of having the measurement event with $ c_k $ photocounts in beam $ k $ for $ k=1,2,3 $. It is important to note that the outcomes in individual double windows are reused multiple times to achieve an effective higher number of measurements, thereby enhancing the experimental precision.

We note that the original neighbor photocount measurements in both the signal and the idler channels exhibit certain weak correlations caused by the correlations among the neighbor pump pulses, as quantified and discussed in Ref.~\cite{PerinaJr.2021}. To eliminate these correlations, we rearranged the original photocount data such that the rearranged neighbor measurements are taken in the detection windows at least $ 10^3 $ measurement realizations apart.
This ensures the independence of individual weak twin beam measurements and allows for the reliable concatenation of the photocount measurements from different weak twin beams.

The analyzed 3-beam Gaussian states differ in the number of weak twin beams used 
per realization with the most complex state incorporating fewer than 200 weak twin beams. Compared to this, the experimental signal and idler photocount channels contain close to $ 7\times 10^8 $ measurements of individual weak twin beams. 
The number of realizations for different 3-beam Gaussian states exceeds $ 10^6 $ provided that the experimental data are not multiply used. 
However, since individual weak twin beam measurements are independent, their multiple use is justified, allowing the number of realizations to increase to approximately $ N_M \approx 7/6\times 10^8 $ (when building the correlated parts of the states, we first join the weak twin beam measurements into groups of six). This improves precision in determining the quantities used, though, for our measurement, the corresponding statistical errors are rather small in both cases.

The experimental histogram $ f(c_1,c_2,c_3) $ allows reconstructing the corresponding photon-number distribution $ p(n_1,n_2,n_3) $ when the detector parameters are known. To simplify the calculations, we assume an effective detector in each beam composed of $ N_{\rm d} = 2w_{\rm p} + w_{\rm n} $ pixels (detection windows) endowed with an average detection efficiency $ \eta = (\eta_{\rm s} + \eta_{\rm i})/2 $ and dark-count rate $ d = (d_{\rm s} + d_{\rm i})(2w_{\rm p} + w_{\rm n}/2) $, where the detection efficiency $ \eta_{\rm s} $ ($ \eta_{\rm i} $) and dark-count rate $ d_{\rm s} $ ($ d_{\rm i} $) characterize the signal (idler) APD. Detection matrix $ T $ of this detector, that gives the probability of having $ c $ photocounts when being illuminated by $ n $ photons, is derived in the following form \cite{PerinaJr2012}:
\begin{eqnarray}     
 T(c,n) &=& \left( \begin{array}{c} N_{\rm d} \cr c \end{array} \right)
  (1-D)^{N_{\rm d}} (1-\eta)^{n} (-1)^{c}  \nonumber \\
  & & \hspace{-10mm} \times \sum_{l=0}^{c}
  \left( \begin{array}{c} c \cr l \end{array} \right) \frac{(-1)^l}{(1-D)^l}
  \left( 1 + \frac{l}{N_{\rm d}} \frac{\eta}{1-\eta}
   \right)^{n} ;
\label{16}
\end{eqnarray}
dark-count rate $ D \equiv d / N _{\rm d}$. Applying the detection matrices of the form (\ref{16}) for all three beams of a symmetric 3-beam Gaussian state, the maximum-likelihood reconstruction method \cite{Dempster1977,Vardi1993} gives us the photon-number distribution $ p(n_1,n_2,n_3) $ as a steady state of the following iteration procedure:
\begin{eqnarray}   
 p^{(j+1)}(n_1,n_2,n_3)&=& p^{(j)}(n_1,n_2,n_3)  \nonumber \\
 & & \hspace{-29mm} \times \sum_{c_1,c_2,c_3=0}^{\infty} F^{(j)}(c_1,c_2,c_3)  T(c_1,n_1)
   T(c_2,n_2) T(c_3,n_3), \nonumber \\
 & &
\label{17} \\
 F^{(j)}(c_1,c_2,c_3) &=& f(c_1,c_2,c_3) \Biggl[
  \sum_{n'_1,n'_2,n'_3=0}^{\infty} T(c_1,n'_1) \nonumber \\
 & & \hspace{-10mm} \times  T(c_2,n'_2) T(c_3,n'_3)
   p^{(j)}(n'_1,n'_2,n'_3) \Bigr]^{-1}, \nonumber \\
 & &   \hspace{2mm} j=0,1, \ldots \; .
 \nonumber
\end{eqnarray}

The photon-number moments $\expect{n_1^{l_1}n_2^{l_2} n_3^{l_3}}$ are then determined along the formula
\begin{eqnarray}   
 \expect{n_1^{l_1}n_2^{l_2} n_3^{l_3}} &=&
  \sum_{n_1,n_2,n_3=0}^{\infty} n_1^{l_1} n_2^{l_2} n_3^{l_3}
  p(n_1,n_2,n_3).
\label{18}
\end{eqnarray} 
The accompanying (integrated) intensity moments $ \expect{W_1^{m_1} W_2^{m_2} W_3^{m_3}} $, that are the normally-ordered photon-number moments, are derived using the Stirling numbers $ S $ of the first kind \cite{Gradshtein2000}:
\begin{eqnarray}   
 \expect{W_1^{m_1} W_2^{m_2} W_3^{m_3}} &=&
  \sum_{l_1=1}^{m_1} S(m_1,l_1) \sum_{l_2=1}^{m_2} S(m_2,l_2)
 \nonumber \\
 & & \hspace{-5mm} \times  \sum_{l_3=1}^{m_3} S(m_3,l_3)
   \langle n_1^{l_1} n_2^{l_2} n_3^{l_3} \rangle.
\label{19}
\end{eqnarray}

The intensity moments of the entire field, denoted as $\expect{W_1^{m_1} W_2^{m_2} W_3^{m_3}}$, need to be reduced to a single effective mode characterized by its intensity moments $\expect{w_1^{m_1} w_2^{m_2} w_3^{m_3}}$. The general method for this reduction is given in Appendix~\ref{AppendixC}. The analyzed states comprise 2 correlated units ($w_{\rm p} = 2$), i.e. 12 double detection windows. According to the analysis presented in Ref.~\cite{PerinaJr.2021}, each weak TWB detected in one detection window comprises 10 independent spatio-spectral modes ($m_{\rm p}=10$). As there are 4 detection windows per beam per one correlated unit, this suggests $M = 80$ effective independent modes ($M = 4w_{\rm p} m_{\rm p}$). However, these independent modes can be grouped together to arrive at effectively more populated modes. Importantly, such grouping results in increased values of the nonclassicality depth \cite{Lee1991}  as well as the entanglement quantifiers \cite{PerinaJr2020a,PerinaJr.2025}. The actual number $ M $ of modes used in the analysis then reflects how the field is detected/monitored and in which way the data are processed. We note that the choice of the number $ M $ of modes determines how we observe the in general rather complex, because multi-mode, field and how we quantify its properties from the chosen point of view. In the following, we provide a detailed analysis of the obtained data assuming $ M = 6.7 $ independent effective modes as this number $ M $ of modes was found suitable when analyzing entanglement and steering in 3-beam symmetric Gaussian states in Ref.°\cite{PerinaJr.2025}.

The reduced intensity moments $\expect{w_1^{m_1} w_2^{m_2} w_3^{m_3}}$ allow us to determine all QUIs $ \Delta^N_k $, $ N=1,2,3$, discussed in the previous section and summarized in Tab.~\ref{tab1}. We have calculated these QUIs for the experimental noisy symmetric 3-beam Gaussian states with the varying mean noise photon number $\expect{n_n}$ per beam. 
The correlated parts of each state consist of 8 detection windows in each beam, which gives $ 0.8527\pm 0.0001 $ mean photon number per beam. The noise parts of the states were made up of 0 to 58 detection windows. Thus, the mean noise photon number $\expect{n_n}$ per beam varies from 0 to 6.3. We use the mean noise photon number $\expect{n_n}$ as an independent variable when drawing the graphs for the QUIs. 

\begin{figure}   
\centering
\includegraphics[width=8cm]{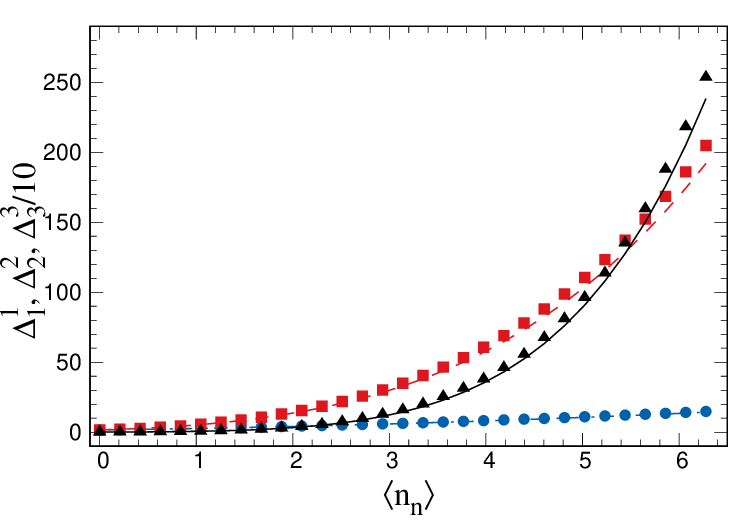}
\caption{Quantum universal invariants per mode $\Delta^1_1$ (blue circles), $\Delta^2_2$ (red squares), and $\Delta^3_3$ (black triangles) as they depend on mean noise photon number $\expect{n_n}$ per beam. 
Isolated symbols describe the experimental data (estimated relative errors better than in turn  1, 2, and 3\% for $\Delta^1_1$, $\Delta^2_2$, and $\Delta^3_3$.), solid curves originate in the Gaussian model presented in Appendix~\ref{AppendixD}.}
\label{fig2}
\end{figure}

The QUIs $\Delta^k_k$ for $k=1,2,3$ uniquely expressible in intensity moments are greater than 1 and increase with the increasing mean noise photon number $\expect{n_n}$, as shown in Fig.~\ref{fig2}.
This reflects the gradual decrease of purities $ \mu_k $, $ k=1,2,3 $, of the symmetric Gaussian states when increasing the mean noise photon number $\expect{n_n}$ as we have $\Delta^k_k = \mu_k^{-1/2}$. We note that the experimental QUIs $\Delta^1_1$ and $\Delta^2_2$ adhere to the known relation $\Delta^2_2 \leq (\Delta^1_1)^2$ (see Eq.~(58) in Ref. \cite{Adesso2004a}).

\begin{figure}   
\centering
\includegraphics[width=8cm]{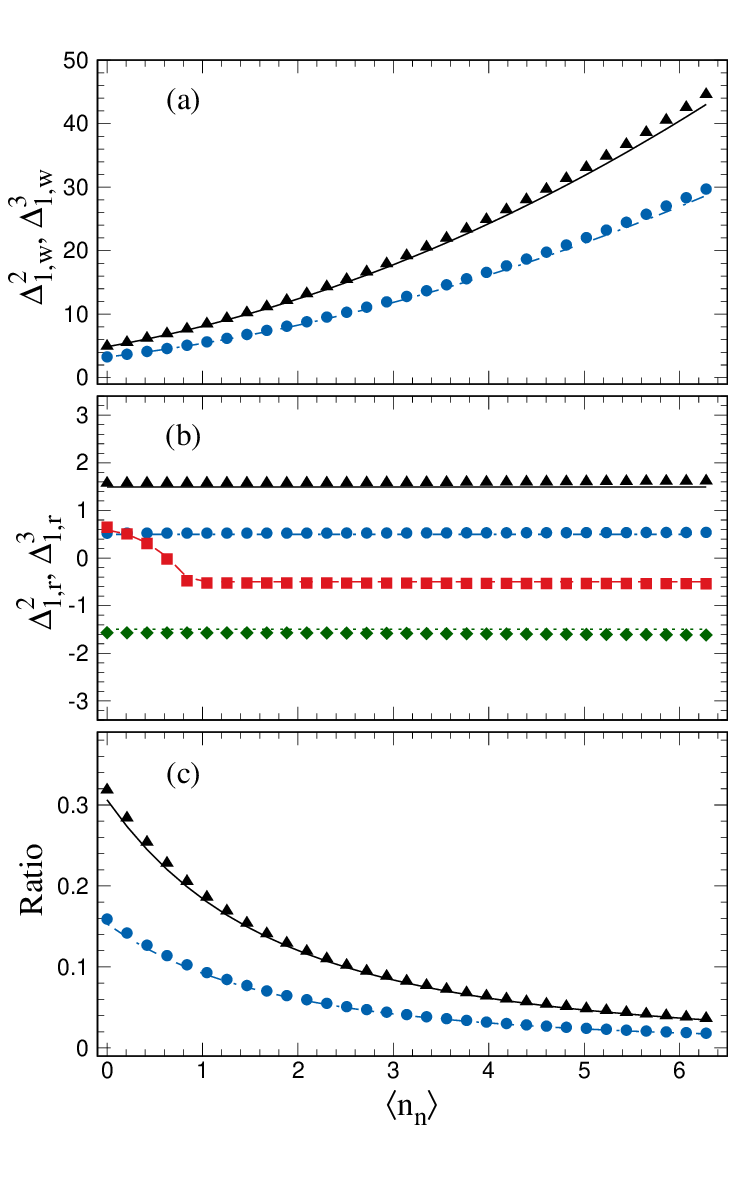} 
\caption{(a) Quantum universal invariants per mode $\Delta^2_1$ and $\Delta^3_1$, expressed via their intensity-moments parts $\Delta^2_{1,w} $ (blue circles) and $\Delta^3_{1,w}$ (black triangles) as they depend on mean noise photon number $\expect{n_n}$ per beam. (b) Residue components of $\Delta^2_{1,r}$ and $\Delta^3_{1_r}$ expressed via their upper 
[$\Delta^2_{1,r}$ (blue circles), $\Delta^3_{1,r} $ (black triangles)] and lower [$\Delta^2_{1,r}$ (red squares), $\Delta^3_{1,r} $ (green diamonds)] bounds given in Eq.~\eqref{eq:boundaries} and Appendix~\ref{AppendixB}. (c) Ratios $ \max\{ | \Delta^2_{1,r}| \}/\Delta^2_{1,w} $ (blue circles) and $ \max\{ |  \Delta^3_{1,r} | \}/\Delta^3_{1,w} $ (black triangles). 
Experimental data are drawn by isolated symbols (relative errors better than 5\% [7\%] for the QUIs and residues [ratios] drawn in (a) and (b) [(c)]); solid curves originate in the Gaussian model presented in Appendix~\ref{AppendixD}. }
\label{fig3}
\end{figure}

The experimental results for the QUIs $\Delta^1_1$, $\Delta^2_2$, and $\Delta^3_3$ drawn in Fig.~\ref{fig2} are compared with the theoretical predictions originating in a multi-mode Gaussian model applied to noisy symmetric 3-beam states (details are described in Appendix~\ref{AppendixD}). The experimental values of QUIs in Fig.~\ref{fig2}, as well as those for the QUIs $\Delta^2_1$, $\Delta^3_1$ and $\Delta^3_2$ plotted in Fig.~\ref{fig3} and \ref{fig4} below, are systematically greater than their theoretical predictions, though the differences are small. This reflects local experimental errors (deviations) in the measurement of the elements in photocount histograms that are kept in the maximum-likelihood reconstruction but are concealed in the fitting of the Gaussian model. Such deviations naturally decrease the state purities $ \mu $ which implies greater values of the QUIs $\Delta^k_k$ for $ k = 1\ldots $. The agreement between the experimental and theoretical values of the QUIs of the symmetric 3-beam states gives the grounds for using these QUIs to characterize the symmetric 3-beam states and their quantum correlations, extending and generalizing the results reached in~\cite{BarasinskiPRL2023} for 2-beam Gaussian states.

\begin{figure}   
\centering
\includegraphics[width=8cm]{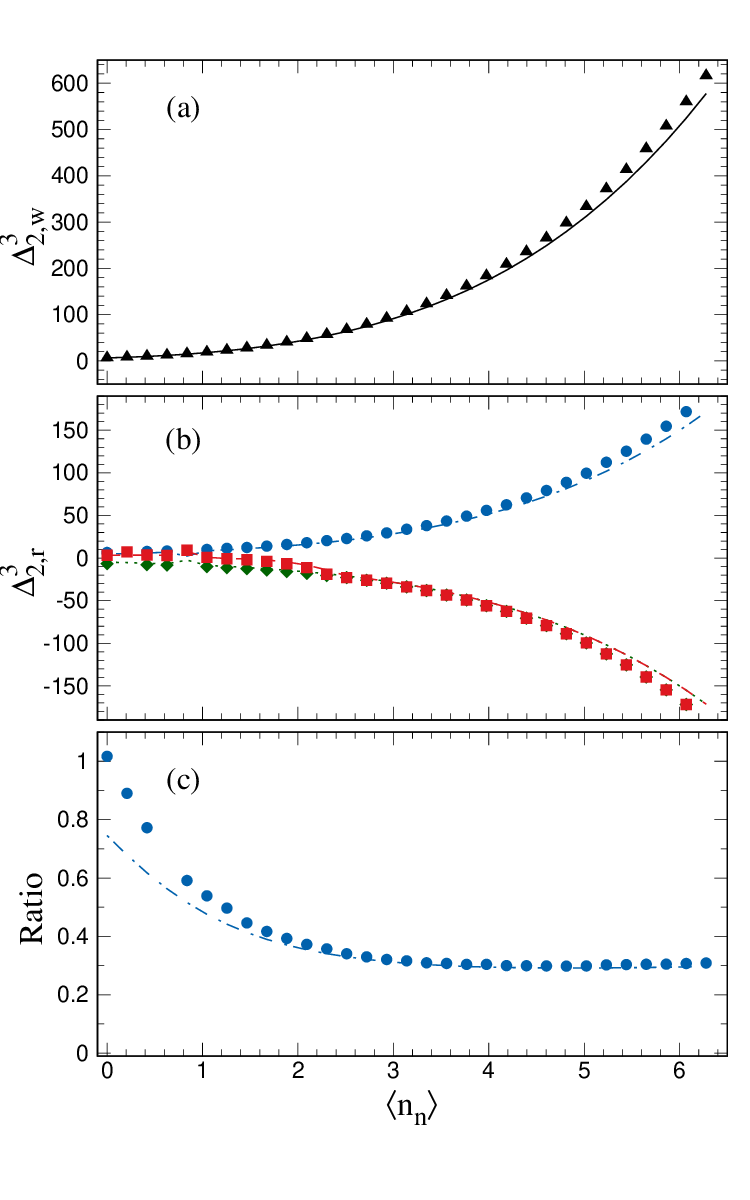} 
\caption{(a) Quantum universal invariant $\Delta^3_2$ per mode expressed via its intensity-moments part $\Delta^3_{2,w}$ (black triangles) as it depends on mean noise photon number $\expect{n_n}$ per beam. 
(b) Residue component $\Delta^3_{2,r}$ characterized by its upper bound (blue circles) and lower bounds in Eq.~\eqref{ResB1n} (green diamonds) and Eq.~\eqref{ResB2n} (red squares). (c) Ratio $ \max\{ | \Delta^3_{2,r} | \}/\Delta^3_{2,w} $ (blue circles). 
Experimental data are drawn by isolated symbols (relative errors better than 5\% [7\%] for the QUIs and residues [ratios] drawn in (a) and (b) [(c)]); solid curves originate in the Gaussian model presented in Appendix~\ref{AppendixD}. }
\label{fig4}
\end{figure}

The remaining QUIs $\Delta^2_{1} $, $\Delta^3_{1} $, and $\Delta^3_{2} $ mentioned in Tab.~\ref{tab1}) cannot be completely expressed in terms of intensity moments. However, their parts $\Delta^2_{1,w} $, $\Delta^3_{1,w} $, and $\Delta^3_{2,w} $ expressible in intensity moments increase with the increasing mean noise photon number $\expect{n_n}$, as documented in Figs.~\ref{fig3}(a) and \ref{fig4}(a). The upper bounds of their residues $\Delta^2_{1,r} $ and $\Delta^3_{1,r}$ (given in Eq.~\eqref{eq:boundaries} and Appendix~\ref{AppendixB}) are practically independent on $\expect{n_n}$, as certified in Fig.~\ref{fig3}(b). Thus, the relative contributions of the residues to these QUIs, quantified by the ratios $ \Delta^2_{1,r}/\Delta^2_{1,w} $ and $ \Delta^3_{1,r}/\Delta^3_{1,w} $, gradually decrease with increasing $\expect{n_n}$ [see Fig.~\ref{fig3}(c)]. They drop below the experimental errors at certain values of the mean noise photon number $\expect{n_n}$. This means that these residues can be neglected for sufficiently large $\expect{n_n} $, and the QUIs $\Delta^2_{1} $ and $\Delta^3_{1} $ are expressible directly along the formulas \eqref{eq:delta21} and \eqref{eq:delta31}, respectively. 
Moreover, when considering both the upper and lower bounds for $\Delta^2_{1,r}$ and $\Delta^3_{1,r} $ (see Eq.~\eqref{eq:boundaries} and Appendix~\ref{AppendixB}), we come to the conclusion that $|\bar{D}{12}| \approx 0$ for $\expect{n_n} < 0.5$. 

On the other hand, the upper bound of the residue $\Delta^3_{2,r}$ increases with increasing $\expect{n_n}$ such that the ratio $ \Delta^3_{2,r}/\Delta^3_{2,w}$ depends only weakly on $\expect{n_n}$ for larger $\expect{n_n}$  (see Fig.~\ref{fig4}(b) and (c)). In this case neglecting $\Delta^3_{2,r}$ could result in significant underestimating $ \Delta^3_{2}$ for small $\expect{n_n}$. Though we note that the constraint in Eq.~\eqref{ResB1n} is quite general and can be improved, for instance, by considering our earlier results regarding $\Delta^3_{1,r}$ or by exploiting the Robertson–Schr\H{o}dinger uncertainty relation outlined in Ref. \cite{SerafiniPRL96_2006}. Using the latter approach, an alternative lower bound for $\Delta^3_{2,r}$ is established [see Eq.~\eqref{ResB2n}]. It clearly demonstrates that the real gap between the upper and lower bounds is much narrower than suggested by Eq.~\eqref{ResB1n}.

We note that the investigated experimental 3-beam Gaussian fields are in a certain sense 'experimental simulations‘ as the data for their individual realizations come from different measurements. This is a consequence of the relatively simple experimental setup based on two APDs used. However, the multi-modality of the investigated fields allows their generation in the form of one pulse, but then real photon-number-resolving detection is needed. An example of such measurement can be found in \cite{Michalek2020} where an iCCD camera was used to characterize the photon-added and photon-subtracted states experimentally. However, such experiments are considerably more complex, less stable, and especially prone to change of experimental conditions when measuring the states with different levels of noise \cite{Thapliyal2024a}.

\section{Identification of entangled and separable states}

The determination of QUIs of the Gaussian states represents an extraordinarily strong tool in the analysis of properties of these states provided that the needed QUIs are determined uniquely using the intensity moments. In this case, knowing the QUIs the state is uniquely identified and its properties are revealed. This is important for applications that require knowledge of the state involving quantum communications and metrology. Symmetric 3-beam Gaussian states are an example of the group of Gaussian states that are uniquely identified by such QUIs. In Ref.~\cite{PerinaJr.2025} we present detailed experimental investigations of the properties of multimode symmetric 3-beam Gaussian states using the approach through QUIs and intensity moments that reveal genuine tripartite entanglement in these states as well as the states that share bipartite and tripartite entanglement. 

\begin{figure}   
\centering
\includegraphics[width=8cm]{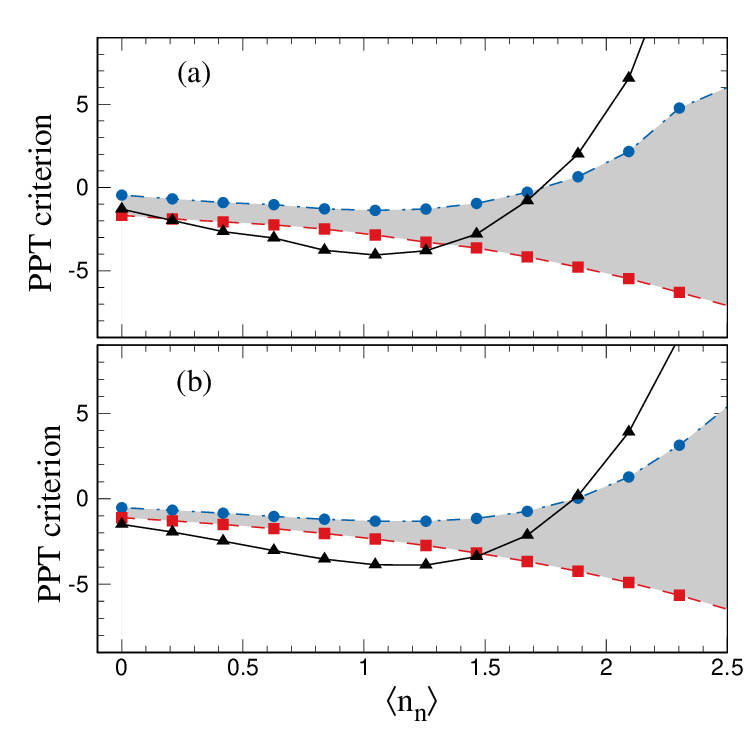} 
\caption{The Robertson-Schrödinger uncertainty relation is analyzed for (a)  experimental symmetric 3-beam  Gaussian states and (b) Gaussian model presented in Appendix~\ref{AppendixD} using partial transposition. Black triangles represent the left-hand side of Eq.~\eqref{PPT_condition}, while the upper and lower bounds of the right-hand side of Eq.~\eqref{PPT_condition} are drawn by blue circles and red squares, respectively.}
\label{fig5}
\end{figure}

If the studied group of Gaussian states is broader and we have to take into account uncertainty in the determination of residues of some QUIs through intensity moments, we can still effectively apply the derived formulas for QUIs and their residues when suitable quantities are under investigation. Such applications can give us important results including the certification about the state entanglement and separability.  
Assume that we detect quantum entanglement through the Peres-Horodecki separability criterion (PPT criterion) \cite{Peresprl77_1996, Horodeckipla232_1997}. This requires the QUIs, $\tilde{\Delta}^n_{k}$, defined for the covariance matrix of a partially transposed state $\tilde{\rho}$.
Introducing the CM $\tilde{{\boldsymbol A}}_{\mathcal{S}}$ of the symmetric 3-beam Gaussian state with partial transposition applied to beam 1, the following relations are derived applying the developed method:
\begin{eqnarray}
    \tilde{\Delta}^3_{1} &=& \Delta^3_{1} + \frac{4}{3} \Delta^3_{1,r},\nonumber\\
    \tilde{\Delta}^3_{2} &=& \Delta^3_{2} + \frac{2}{3} \Delta^3_{2,r}- (|D_{12}|^2+|\bar{D}_{12}|^2),\nonumber\\   
    \tilde{\Delta}^3_{3} &=& \Delta^3_{3}.\nonumber
\end{eqnarray}
The necessary condition for separability derived from PPT criterion gives us the inequality $\tilde{\Delta}^3_{3} - \tilde{\Delta}^3_{2} + \tilde{\Delta}^3_{1} - 1 \geq 0$ (for details, see Ref. \cite{SerafiniPRL96_2006}). This condition also certifies that $\tilde{{\boldsymbol A}}_{\mathcal{S}}$ satisfies the Robertson-Schr\"{o}dinger uncertainty relation. Violation of this inequality means that the analyzed state is entangled. 
The derived uncertainty relation can simply be reformulated
\begin{eqnarray}   
    \Delta^3_{3} - \Delta^3_{2,w} &+& \Delta^3_{1,w} - 1 + (|D_{12}|^2+|\bar{D}_{12}|^2) \geq\nonumber\\
    && -\frac{1}{3} ({\Delta}^3_{2,r} - {\Delta}^3_{1,r}),
    \label{PPT_condition}
\end{eqnarray}
where ${\Delta}^3_{2,r}$ and ${\Delta}^3_{1,r}$ obey the limitations given in Eq.~\eqref{eq:boundaries} and Appendix~\ref{AppendixB}. Applying the inequality \eqref{PPT_condition} to the experimentally generated symmetric 3-beam Gaussian states in Fig. \ref{fig5} (a) we reveal that these states are entangled for $\expect{n_n} < 1.3$ and separable (possibly exhibiting bound entanglement) for $\expect{n_n} > 1.8$. Uncertainty in the determination of residues $ {\Delta}^3_{2,r} $ and $ {\Delta}^3_{1,r}$ does not allow to decide about the entanglement for
$\expect{n_n} \in (1.3,1.8) $. Similar conclusions are drawn for the accompanying Gaussian model presented in Appendix D using Fig. \ref{fig5} (b).

\section{Conclusions}

A general method has been developed to express the quantum universal invariants of an N-beam Gaussian field in terms of experimental intensity moments. While some invariants can be fully determined by intensity moments, others include residual terms that cannot be directly accessed through these measurements. However, these residuals can be bounded using expressions based on intensity moments, which allows estimation of their contribution.

This method has been applied to 1-, 2-, and 3-beam Gaussian fields, giving explicit formulas for the corresponding invariants. These formulas, as an example,  were applied to analyze the class of symmetric 3-beam Gaussian fields characterized by their mean noise photon numbers. For sufficiently large noise photon numbers, the residual contributions to the invariants can be neglected. When the values of invariants are known, different forms of quantum correlations are directly quantified. In particular, the Peres-Horodecki separability criterion was formulated in terms of the experimental intensity moments to decide about entanglement or separability of the state. This approach eliminates the need for complete knowledge of the symplectic eigenvalues of the partially transposed covariance matrix, significantly simplifying its application to general Gaussian fields.

The presented method can be readily extended to arbitrary multipartite Gaussian states to derive their quantum universal invariants. This may be helpful in reconstructing the Gaussian states, particularly in their Williamson normal form, using photon-number measurements.

Given the widespread application of the Gaussian states in metrology and in general, in quantum information processing including quantum communications, this method holds significant promise for advancing these fields in the near future.

Data underlying the results presented in this work may be obtained from the  repository \cite{zenodo}.

\acknowledgments

J.P. acknowledges support by the project ITI  CZ.02.01.01/00/23\_021/0008790 of the Ministry of Education, Youth, and Sports of the Czech Republic and EU.

\appendix

\section{Quantum Universal Invariants for 3-Beam Gaussian states}
\label{AppendixA}

Below we present the QUIs for a general
3-beam Gaussian field expressed in terms of intensity moments. The
first QUI $ \Delta^3_1 $ is given as
\begin{eqnarray}
    \Delta^3_1 &=& 3 + \{ 4 \expect{W_1} + 12 \expect{W_1}^2 - 4 \expect{W_1^2}
    + {\rm addt}_1 \} \nonumber\\
    & & -  \Delta^3_{1,r} ,
    \label{eq:delta31}
\end{eqnarray}
where $ \Delta^3_{1,r} = \{ 4(|D_{12}|^2-|\bar{D}_{12}|^2) + {\rm
addt}_2 \} $. Symbol $ {\rm addt}_1 $ replaces the terms that are
obtained by substituting index 1 by indices 2 and 3. Similarly,
symbol $ {\rm addt}_2 $ replaces the terms that are obtained by
substituting indices \{1,2\} by indices \{2,1\},\{1,3\},\{3,1\},
\{2,3\}, and \{3,2\} .

The second QUI $ \Delta^3_2 $ is derived in the form: 
\begin{widetext}
\begin{eqnarray}
 \Delta^3_2 &=&3 + \{8 \expect{W_1} -8\expect{W_1^2} + 24 \expect{W_1}^2 + {\rm add}_1 \} +
   \{  - 4 \expect{W_1^2W_2^2} + 8 \expect{W_1W_2^2}+ 24 \expect{W_1W_2}^2
   \nonumber \\
  & &   + 8 \expect{W_1} \expect{W_2}  +  12 \expect{W_1^2}\expect{W_2^2} - 24 \expect{W_1^2}\expect{W_2} - 48 \expect{W_1} \expect{W_1W_2}
   + 48 \expect{W_1}\expect{W_1W_2^2}    + 96 \expect{W_1}^2 \expect{W_2}
     \nonumber \\
  & & - 96 \expect{W_1^2} \expect{W_2}^2 + 240 \expect{W_1}^2
  \expect{W_2}^2 - 192 \expect{W_1} \expect{W_1W_2} \expect{W_2}
     + {\rm addt}_2 \}- \Delta^3_{2,r},
 \label{eq:delta32}
\end{eqnarray}
where
\begin{eqnarray}
 \Delta^3_{2,r} &=& \{ 8(| D_{12} |^2 + {\rm addt}_2 \} + \{ 16 (B_1 + B_1^2 - |C_1|^2)
  (| D_{23} |^2 - | \bar{D}_{23} |^2)
  + 16 \Re \{D_{12} D_{13}^{\ast} \bar{D}_{23}\} + 32(B_1 + B_2 - B_3)
  \nonumber \\
  & & \times \Re \{\bar{D}_{12} D_{23}^{\ast} D_{13}\} - 16(2B_1+1) \Re \{\bar{D}_{12} \bar{D}_{13}^{\ast} \bar{D}_{23}\}
  -16( |D_{12} |^2  - |\bar{D}_{12} |^2)( | D_{13} |^2 - | \bar{D}_{13} |^2 )   \nonumber \\
  & &  +32( \Re\{C_1 D_{12}^*D_{13}^*D_{23}\} + \Re\{C_1\bar D_{12}\bar D_{13}D_{23}^{\ast}\} )
   -64\Re\{C_2\bar D_{13}\bar D_{12}^{\ast}D_{23}^{\ast}\} + {\rm addt}_3 \}.
 \label{eq:R32}
\end{eqnarray}
Symbol $ {\rm addt}_3 $ replaces the terms that are obtained by
substituting indices \{1,2,3\} by indices
\{1,3,2\},\{2,1,3\},\{3,1,2\}, \{2,3,1\}, and \{3,2,1\}. We note
that $ D_{jk} = D_{kj} $ and $ \bar{D}_{jk} = \bar{D}_{kj}^* $ for
$ j \neq k $ . 

Finally, we attain the third QUI $ \Delta^3_3 $ in the form: 
\end{widetext} 

\begin{widetext}
\begin{eqnarray}
    \Delta^3_3 &=& 1 + \{ 4 \expect{W_1}  + 4 \expect{W_1^2} + 12 \expect{W_1}^2
    + {\rm addt}_1 \} + \{ - 4\expect{W_1W_2} - 4 \expect{W_1^2W_2^2} + 8 \expect{W_1W_2^2} + 24 \expect{W_1W_2}^2  \nonumber \\
     & & + 12 \expect{W_1} \expect{W_2} + 12 \expect{W_1^2} \expect{W_2^2} - 24 \expect{W_1^2} \expect{W_2}
      - 96 \expect{W_1^2}\expect{W_2}^2  + 96 \expect{W_1}^2\expect{W_2}
     + 240 \expect{W_1}^2 \expect{W_2}^2 \nonumber \\
    & &  - 48 \expect{W_1} \expect{W_1W_2} + 48 \expect{W_1} \expect{W_1W_2^2}
     - 192 \expect{W_1} \expect{W_1W_2} \expect{W_2}  + {\rm addt}_2 \} \nonumber \\
    & & + \{ 8/3 \expect{W_1W_2W_3}  - 8/3\expect{W_1^2W_2^2W_3^2} - 8 \expect{W_1W_2W_3^2} + 8 \expect{W_1W_2^2W_3^2} - 24 \expect{W_1W_3}\expect{W_2} - 24 \expect{W_1^2W_3^2} \expect{W_2} \nonumber \\
    & &  +24\expect{W_1^2}\expect{W_2W_3} + 24\expect{W_1^2W_3^2}\expect{W_2^2} + 32 \expect{W_1W_2W_3}^2
     + 32 \expect{W_1} \expect{W_2} \expect{W_3} - 32 \expect{W_1^2} \expect{W_2^2} \expect{W_3^2} \nonumber \\
    & &  + 48 \expect{W_1} \expect{W_1W_2W_3} + 48\expect{W_1}\expect{W_1W_2^2W_3^2} +48
    \expect{W_1W_3^2}\expect{W_2} - 48 \expect{W_1^2}
    \expect{W_2W_3^2} + 48 \expect{W_1W_2} \expect{W_1W_3} \nonumber \\
    & &   + 48 \expect{W_1W_2^2}\expect{W_1W_3^2} - 96 \expect{W_1W_3} \expect{W_2}^2  -96 \expect{W_1W_2} \expect{W_1W_2W_3}  -  96 \expect{W_1} \expect{W_1W_2W_3^2} \nonumber \\
    & &  + 96 \expect{W_1W_2} \expect{W_1W_2W_3^2}
     - 96 \expect{W_1^2W_3^2} \expect{W_2}^2   - 96 \expect{W_1W_2^2}\expect{W_1W_3}
     + 192 \expect{W_1W_3}^2 \expect{W_2} +192\expect{W_1W_3^2}\expect{W_2}^2 \nonumber \\
    & & - 192 \expect{W_1^2}\expect{W_2W_3}^2  + 960 \expect{W_1W_3}^2\expect{W_2}^2 - 96\expect{W_1^2} \expect{W_2} \expect{W_3} + 96 \expect{W_1^2} \expect{W_2^2} \expect{W_3}     \nonumber \\
    & &  + 384 \expect{W_1}\expect{W_1W_3^2} \expect{W_2} + 384\expect{W_1} \expect{W_1W_2W_3} \expect{W_2} - 384 \expect{W_1} \expect{W_1W_2W_3^2} \expect{W_2} - 384 \expect{W_1} \expect{W_1W_3} \expect{W_2}
      \nonumber \\
    & & +384 \expect{W_1^2} \expect{W_2} \expect{W_2W_3} - 384 \expect{W_1^2} \expect{W_2}\expect{W_2W_3^2}
        +  480 \expect{W_1}^2 \expect{W_2} \expect{W_3} + 480 \expect{W_1}^2 \expect{W_2^2} \expect{W_3^2}  \nonumber \\
    & &  + 768 \expect{W_1W_2} \expect{W_1W_3} \expect{W_2}  - 768 \expect{W_1W_2W_3} \expect{W_1W_3} \expect{W_2} - 768 \expect{W_1W_2} \expect{W_1W_3^2}\expect{W_2}  \nonumber\\
    & &  - 768 \expect{W_1W_2} \expect{W_1W_3} \expect{W_2W_3}  - 960 \expect{W_1} \expect{W_2}^2 \expect{W_3^2} - 1920 \expect{W_1} \expect{W_1W_3} \expect{W_2}^2  + 1920 \expect{W_1} \expect{W_1W_3^2} \expect{W_2}^2
        \nonumber\\
    & &  + 2880\expect{W_1}^2 \expect{W_2} \expect{W_3}^2  + 6720 \expect{W_1}^2 \expect{W_2}^2 \expect{W_3}^2  - 2880 \expect{W_1}^2 \expect{W_2}^2 \expect{W_3^2}
     + 1280 \expect{W_1} \expect{W_1W_2W_3} \expect{W_2} \expect{W_3}  \nonumber\\
      & & - 1920\expect{W_1} \expect{W_1W_2}\expect{W_2}\expect{W_3} + 1920 \expect{W_1^2} \expect{W_2} \expect{W_2W_3} \expect{W_3} +3840 \expect{W_1} \expect{W_1W_3} \expect{W_2} \expect{W_2W_3}
          \nonumber\\
    & & - 11520 \expect{W_1} \expect{W_1W_3} \expect{W_2}^2 \expect{W_3}     + {\rm addt}_3 \}.
    \label{eq:delta33}
\end{eqnarray}
\end{widetext}

\section{Simplified formulas for residue of invariant $\Delta_2^3$}
\label{AppendixB}

When a symmetric 3-beam Gaussian state is considered, the formula for residue of the QUI $\Delta_2^3$ is derived in the following form 
\begin{eqnarray}   
\Delta^3_{2,r}  &=&48|D_{12}|^2\nonumber\\
&+&96(B_1+B_1^2-|C_1|^2)(|D_{12}|^2-|\bar D_{12}|^2)\nonumber\\
&-&96(|D_{12}|^2-|\bar D_{12}|^2)^2\nonumber\\
&+&192 B_1\Re\{\bar D_{12}\}(|D_{12}|^2-|\bar D_{12}|^2)\nonumber\\
&+&96 \Re\{\bar D_{12}\}(|D_{12}|^2-|\bar D_{12}|^2)\nonumber\\
&+&192 \Re\{C_1 D^*_{12}\}(|D_{12}|^2-|\bar D_{12}|^2)
\label{ResB1}
\end{eqnarray}
that admits the following estimates.

The upper and lower bounds of the residue in Eq. \eqref{ResB1} can be obtained using the inequality $|x+y| \le |x|+|y|$, which enables to evaluate each term separately. Then, applying the inequalities \eqref{eq:res1} and Eq. \eqref{eq:intensity}, straightforward calculations leave us with the inequality
\begin{eqnarray}
|\Delta^3_{2,r}|  &\le &48 (\expect{W_1W_2}- \expect{W_1}\expect{W_2})\nonumber\\
&+&96|\expect{W_1} + 3 \expect{W_1}^2 - \expect{W_1^2}| 
 (\expect{W_1W_2}- \expect{W_1}\expect{W_2})\nonumber\\
&+&96(\expect{W_1W_2}- \expect{W_1}\expect{W_2})^2\nonumber\\
&+&96\expect{W_1} \Big|\expect{W_1}^3+3 \expect{W_1} \expect{W_1W_2} - \expect{W_1W_2W_3}\nonumber\\
&-&3 \expect{W_1}^2 \expect{W_2}\Big|\nonumber\\
&+&48 \Big|\expect{W_1}^3+3 \expect{W_1} \expect{W_1W_2} 
-\expect{W_1W_2W_3}\nonumber\\
&-& 3 \expect{W_1}^2 \expect{W_2}\Big|\nonumber\\
&+&48 \Big|(4 \expect{W_1^2 W_2W_3} - 8 \expect{W_1W_2} \expect{W_1W_3} \nonumber\\
&-& \expect{W_1^2 W_3} \expect{W_2} - 2 \expect{W_1^2} \expect{W_2W_3} - \expect{W_1^2 W_2} \expect{W_3} \nonumber\\
&+& 2 \expect{W_1^2} \expect{W_2} \expect{W_3} + 4 \expect{W_1} (2 \expect{W_1}^3 \nonumber\\
&+& 6 \expect{W_1} \expect{W_1W_2} - 2 \expect{W_1W_2W_3} + \expect{W_1W_3} \expect{W_2} \nonumber\\
&-& 6 \expect{W_1}^2 \expect{W_2} + \expect{W_1W_2} \expect{W_3}))\Big| \nonumber\\
&+& 48 \sqrt{\expect{W_1W_2} - \expect{W_1} \expect{W_2}} \Big|-\expect{W_1^2W_2} \nonumber\\
&+& 4 \expect{W_1} \expect{W_1W_2} - 4 \expect{W_1}^2 \expect{W_2} + \expect{W_1^2} \expect{W_2}\Big|.
\label{ResB1n}
\end{eqnarray}
To express the last line of  Eq.~\eqref{ResB1} we apply the following approach. We define $A \equiv \Re\{C_1 D^*_{12}\}(|D_{12}|^2-|\bar D_{12}|^2)$ and $B\equiv 2 \Re\{\bar{D}_{12}\} \Re\{C_1 D^*_{12} \bar{D}_{12}\}$ and then we rely on the following inequality 
\begin{eqnarray}
|A| = |A+B-B|\leq |A+B| + |B|.\nonumber
\end{eqnarray}
We have $A+B= 2 \Re\{C_1 D^*_{12} \bar{D}^2_{12}\}+ \Re\{C_1 D^*_{12}\}(|D_{12}|^2+|\bar D_{12}|^2)$ and so we can express it using intensity moments up to $\expect{W^2_1W_2W_3}$. On the other hand, $|B|=2 | \Re\{\bar{D}_{12}\}|~|\Re\{C_1 D^*_{12} \bar{D}_{12}\}|\le (|D_{12}|^2+|\bar{D}_{12}|^2)^{1/2}~|\Re\{C_1 D^*_{12} \bar{D}_{12}\}|$. 

Now we consider the relation $\Delta^3_{3}-\Delta^3_{2}+\Delta^3_{1}-1\geq 0$ derived in Ref.~\cite{SerafiniPRL96_2006}. Substituting $\Delta^3_{2} = \Delta^3_{2,w}-\Delta^3_{2,r}$ and $\Delta^3_{1} = \Delta^3_{1,w}-\Delta^3_{1,r}$ we arrive at
\begin{eqnarray}
    \Delta^3_{2,r} &\geq& -\Delta^3_{3}+\Delta^3_{2,w}-\Delta^3_{1,w}+\Delta^3_{1,r}+1 \nonumber\\
    &=& -\Delta^3_{3}+\Delta^3_{2,w}-\Delta^3_{1,w}+3 \Delta^2_{1,r}+1.
\end{eqnarray}
The last equality comes from the block structure of the covariance matrix, where $\Delta^3_{1,r}$ is expressed as the sum of $\Delta^2_{1,r}$ terms calculated for  three two-beam subsystems. For a symmetric 3-beam Gaussian state $\Delta^2_{1,r}$ is the same for all two-beam subsystems and it can be estimated by the relation \eqref{eq:boundaries}. 
\begin{eqnarray}
\label{ResB2n}
    \Delta^3_{2,r} &\geq& -\Delta^3_{3}+\Delta^3_{2,w}-\Delta^3_{1,w}+1\\
    & & \hspace{-5mm} + 3\max\{-\Delta^2_2 + \Delta^2_{1,w}-1 ,  -8\expect{W_1W_2}+8\expect{W_1} \expect{W_2} \}.\nonumber
\end{eqnarray}
The residue $\Delta^3_{2,r}$ has to satisfy both relations written in Eqs.~\eqref{ResB1n} and \eqref{ResB2n}.

\section{Relations among single-mode and multi-mode intensity moments}
\label{AppendixC}

We consider an $ M $ mode field present in all three beams.
Labeling independent modes with identical properties by index
$ m $, $ m =1, \ldots, M $, the mean intensity $ \langle W_j \rangle $
of beam $ j $ is given as $ \sum_{m=1}^{M} \langle w_{j,m} \rangle
= M \langle w_j \rangle $. Assuming the Gaussian character of the
analyzed fields, we may establish relations among the moments of
the intensity fluctuations $ \Delta W_j = W_j-\langle W_j\rangle $
and $ \Delta w_j = w_j-\langle w_j\rangle $, $ j=1,2,3 $
\cite{Perina1991,Mandel1995}. First, we consider the moments of
individual beams:
\begin{eqnarray}   
 \langle W_j\rangle &=& M \langle w_j\rangle, \nonumber \\
 \langle (\Delta W_j)^2 \rangle &=& M \langle (\Delta w_j)^2 \rangle, \nonumber \\
 \langle (\Delta W_j)^3 \rangle &=& M \langle (\Delta w_j)^3
 \rangle,  \hspace{1mm} j=1,2,3.
\label{C1}
\end{eqnarray}
The moments encompassing intensities of two beams obey the relations:
\begin{eqnarray}   
 \langle \Delta W_j \Delta W_k \rangle &=& M \langle \Delta w_j \Delta w_k \rangle, \nonumber \\
 \langle (\Delta W_j)^2 \Delta W_k \rangle &=& M \langle (\Delta w_j)^2 \Delta w_k \rangle, \nonumber \\
 \langle (\Delta W_j)^2 (\Delta W_k)^2 \rangle &=& M \langle (\Delta w_j)^2 (\Delta w_k)^2 \rangle \nonumber \\
   & & \hspace{-25mm} + M(M-1) [ 2\langle \Delta w_j \Delta w_k \rangle^2
    + \langle (\Delta w_j)^2\rangle \langle (\Delta w_k)^2 \rangle ], \nonumber \\
 \langle (\Delta W_j)^3 \Delta W_k \rangle &=& M \langle (\Delta w_j)^3 \Delta w_k \rangle  + 3M(M-1) \nonumber \\
   & & \hspace{-25mm} \times \langle (\Delta w_j)^2\rangle \langle \Delta w_j \Delta w_k \rangle, \nonumber \\
 \langle (\Delta W_j)^3 (\Delta W_k)^2 \rangle &=& M \langle (\Delta w_j)^3 (\Delta w_k)^2 \rangle + M(M-1) \nonumber \\
   & & \hspace{-25mm} \times [3 \langle (\Delta w_j)^2\rangle \langle \Delta w_j (\Delta w_k)^2 \rangle
    + \langle (\Delta w_j)^3\rangle \langle (\Delta w_k)^2 \rangle , \nonumber \\
   & & \hspace{-25mm} + 6\langle (\Delta w_j)^2\Delta w_k\rangle \langle \Delta w_j \Delta w_k \rangle ], \nonumber \\
 \langle (\Delta W_j)^3 (\Delta W_k)^3 \rangle &=& M \langle (\Delta w_j)^3 (\Delta w_k)^3 \rangle + M(M-1)\nonumber \\
   & & \hspace{-25mm} \times [3 \langle (\Delta w_j)^2\rangle \langle \Delta w_j (\Delta w_k)^3 \rangle
    + \langle (\Delta w_j)^3\rangle \langle (\Delta w_k)^3 \rangle \nonumber \\
   & & \hspace{-25mm} + 3 \langle (\Delta w_j)^3\Delta w_k\rangle \langle (\Delta w_k)^2 \rangle   + 9\langle (\Delta w_j)^2(\Delta w_k)^2\rangle \nonumber \\
   & & \hspace{-25mm}  \times \langle \Delta w_j\Delta w_k\rangle ] 
     + M(M-1)(M-2) [9\langle (\Delta w_j)^2\rangle \nonumber \\
   & & \hspace{-25mm}  \times  \langle \Delta w_j\Delta w_k \rangle \langle (\Delta w_k)^2\rangle
     + 6\langle \Delta w_j\Delta w_k \rangle^3 ]. \nonumber \\
  & &  \hspace{1mm} j,k=1,2,3, j\ne k .
\label{C2}
\end{eqnarray}
The moments containing intensities of all three beams fulfill the
following relations:
\begin{eqnarray}   
 \langle \Delta W_j \Delta W_k \Delta W_l \rangle &=& M \langle \Delta w_j \Delta w_k \Delta w_l \rangle, \nonumber \\
 \langle (\Delta W_j)^2 \Delta W_k \Delta W_l \rangle &=& M \langle (\Delta w_j)^2 \Delta w_k \Delta w_l \rangle
    \nonumber \\
   & & \hspace{-30mm} + M(M-1) [\langle (\Delta w_j)^2\rangle \langle \Delta w_k \Delta w_l \rangle     \nonumber \\
   & & \hspace{-30mm} +
    2\langle \Delta w_j\Delta w_k\rangle \langle \Delta w_j \Delta w_l\rangle], \nonumber \\
 \langle (\Delta W_j)^2 (\Delta W_k)^2 \Delta W_l \rangle &=& M \langle (\Delta w_j)^2 (\Delta w_k)^2 \Delta w_l \rangle
    \nonumber \\
   & & \hspace{-40mm} + M(M-1) [\langle (\Delta w_j)^2\rangle \langle (\Delta w_k)^2 \Delta w_l \rangle \nonumber \\
   & & \hspace{-40mm} + \langle (\Delta w_k)^2\rangle \langle (\Delta w_j)^2 \Delta w_l \rangle\nonumber \\
   & & \hspace{-40mm}  +  4\langle \Delta w_j\Delta w_k\rangle \langle \Delta w_j \Delta w_k \Delta w_l\rangle \nonumber \\
    & & \hspace{-40mm} + 2\langle \Delta w_j\Delta w_l\rangle \langle \Delta w_j (\Delta  w_k)^2\rangle\nonumber \\
   & & \hspace{-40mm} + 2 \langle \Delta w_k\Delta w_l\rangle \langle (\Delta w_j)^2 \Delta w_k \rangle,
    \nonumber \\
  \langle (\Delta W_j)^2 (\Delta W_k)^2 (\Delta W_l)^2 \rangle &=& M \langle (\Delta w_j)^2 (\Delta w_k)^2 (\Delta w_l)^2 \rangle
    \nonumber \\
   & & \hspace{-40mm} + M(M-1) [\langle (\Delta w_j)^2\rangle \langle (\Delta w_k)^2 (\Delta w_l)^2 \rangle \nonumber \\
    & & \hspace{-40mm} + \langle (\Delta w_k)^2\rangle \langle (\Delta w_j)^2 (\Delta w_l)^2 \rangle\nonumber \\
   & & \hspace{-40mm} + \langle (\Delta w_l)^2\rangle \langle (\Delta w_j)^2 (\Delta w_k)^2 \rangle \nonumber \\
    & & \hspace{-40mm} + 4\langle \Delta w_j\Delta w_k\rangle \langle \Delta w_j \Delta w_k (\Delta w_l)^2\rangle  \nonumber \\
    & & \hspace{-40mm}  + 4\langle \Delta w_j\Delta w_l\rangle \langle \Delta w_j (\Delta w_k)^2 \Delta w_l \rangle \nonumber \\
    & & \hspace{-40mm} + 4\langle \Delta w_k\Delta w_l\rangle \langle (\Delta w_j)^2 \Delta w_k \Delta w_l\rangle +
    8\langle \Delta w_j\Delta w_k\Delta w_l\rangle^2  \nonumber \\
   & & \hspace{-40mm} + M(M-1)(M-2) [\langle (\Delta w_j)^2\rangle \langle (\Delta w_k)^2\rangle \langle (\Delta w_l)^2\rangle
     \nonumber \\
    & & \hspace{-40mm}  + 4 \langle \Delta w_j\Delta w_k\rangle^2 \langle (\Delta w_l)^2\rangle
    + 4 \langle \Delta w_j\Delta w_l\rangle^2 \langle (\Delta w_k)^2\rangle \nonumber \\
    & & \hspace{-40mm} + 4 \langle \Delta w_k\Delta w_l\rangle^2 \langle (\Delta w_j)^2\rangle \nonumber \\
    & & \hspace{-40mm} + 8 \langle\Delta w_j\Delta w_k\rangle \langle \Delta w_k\Delta w_l\rangle
     \langle \Delta w_j\Delta w_l\rangle ], \nonumber \\
  \langle (\Delta W_j)^3 \Delta W_k \Delta W_l \rangle &=& M \langle (\Delta w_j)^3 \Delta w_k \Delta w_l \rangle
    \nonumber \\
   & & \hspace{-40mm} + M(M-1) [\langle (\Delta w_j)^3\rangle \langle \Delta w_k \Delta w_l \rangle\nonumber \\
    & & \hspace{-40mm} + 3\langle (\Delta w_j)^2\rangle \langle \Delta w_j\Delta w_k \Delta w_l \rangle\nonumber \\
    & & \hspace{-40mm} + 3\langle \Delta w_j\Delta w_k \rangle \langle (\Delta w_j)^2\Delta w_l \rangle\nonumber \\
    & & \hspace{-40mm} + 3\langle \Delta w_j\Delta w_l \rangle \langle (\Delta w_j)^2\Delta w_k\rangle], \nonumber \\
  \langle (\Delta W_j)^3 (\Delta W_k)^2 \Delta W_l \rangle &=& M \langle (\Delta w_j)^3 (\Delta w_k)^2 \Delta w_l \rangle
    \nonumber \\
   & & \hspace{-40mm} + M(M-1) [3\langle (\Delta w_j)^2\rangle \langle \Delta w_j (\Delta w_k)^2 \Delta w_l \rangle\nonumber \\
   & & \hspace{-40mm}  + \langle (\Delta w_k)^2\rangle \langle (\Delta w_j)^3  \Delta w_l
   \rangle\nonumber \\
   & & \hspace{-40mm} + 6 \langle \Delta w_j\Delta w_k\rangle \langle (\Delta w_j)^2 \Delta w_k\Delta w_l \rangle\nonumber \\
   & & \hspace{-40mm} + 3 \langle \Delta w_j\Delta w_l\rangle \langle (\Delta w_j)^2 (\Delta w_k)^2 \rangle\nonumber \\
   & & \hspace{-40mm} + 2 \langle \Delta w_k\Delta w_l\rangle \langle (\Delta w_j)^3 \Delta w_k \rangle\nonumber \\
   & & \hspace{-40mm} + \langle (\Delta w_j)^3\rangle \langle (\Delta w_k)^2 \Delta w_l \rangle\nonumber \\
   & & \hspace{-40mm} + 6 \langle (\Delta w_j)^2\Delta w_k\rangle \langle \Delta w_j \Delta w_k \Delta w_l \rangle\nonumber \\
   & & \hspace{-40mm}  + 3 \langle (\Delta w_j)^2\Delta w_l\rangle \langle \Delta w_j (\Delta w_k)^2\rangle\nonumber \\
   & & \hspace{-40mm} + M(M-1)(M-2) [3\langle (\Delta w_j)^2\rangle \langle (\Delta w_k)^2\rangle \langle \Delta w_j\Delta w_l \rangle \nonumber \\
   & & \hspace{-40mm} + 6\langle \Delta w_j^2\rangle \langle \Delta w_j\Delta w_k\rangle 
    \langle \Delta w_k\Delta w_l\rangle \nonumber \\
   & & \hspace{-40mm} + 6\langle \Delta w_j\Delta w_k\rangle^2 \langle \Delta w_j\Delta
    w_l\rangle]\nonumber \\
  & &  \hspace{-20mm} j,k,l=1,2,3, j\ne k\ne l \ne j.
\label{C3}
\end{eqnarray}

Eqs.~(\ref{C1}), (\ref{C2}) and (\ref{C3}) give linear relations
between the moments $ \langle (\Delta W_1)^{k_1} (\Delta
W_2)^{k_2} (\Delta W_3)^{k_3}\rangle $ of the intensity fluctuations of
the whole field and the moments $ \langle (\Delta w_1)^{k_1}
(\Delta w_2)^{k_2} (\Delta w_3)^{k_3}\rangle $ of intensity fluctuations
in one mode. Their recursive character allows, using suitable
order of their solution, to express the moments $ \langle (\Delta
w_1)^{k_1} (\Delta w_2)^{k_2} (\Delta w_3)^{k_3}\rangle $ of one mode in
terms of the moments $ \langle (\Delta W_1)^{k_1} (\Delta
W_2)^{k_2} (\Delta W_3)^{k_3}\rangle $ of the whole field once we know
the number $ M $ of modes.

\section{Gaussian model of 3-mode symmetric Gaussian states}
\label{AppendixD}
For comparison, we have approximated a weak TWB detected in one
detection double window by a multi-mode Gaussian noisy TWB with
suitable parameters and used sequences of such TWBs to model the properties of
3-mode symmetric Gaussian states with their structure
outlined in Sec.~\ref{Experiment}.

Using the method of detector calibration \cite{PerinaJr2012a}, we
have estimated parameters of a weak TWB detected by two APDs and also
determined detection efficiencies $ \eta_{\rm s} $ and $ \eta_{\rm
i} $ of the APDs. In accordance with the properties of spontaneous
parametric down-conversion we assume the weak TWB composed of the
paired-photon, noise signal-photon, and noise idler-photon parts
with the corresponding photon-number distributions $ p_{\rm p} $,
$ p_{\rm s} $, and $ p_{\rm i} $. They take the form of a
multi-mode thermal Mandel-Rice distribution \cite{Perina1991} that
characterizes a field with $ m_a $ equally populated modes each
having mean photon (-pair) number $ b_a $:
\begin{equation}  
  p_a(n;m_a,b_a) = \frac{\Gamma(n+m_a) }{n!\, \Gamma(m_a)}
  \frac{b_a^n}{(1+b_a)^{n+b_a}}, \hspace{5mm} a={\rm p,s,i}
\label{D1}
\end{equation}
and $ \Gamma $ denotes the $ \Gamma $-function. The joint
photon-number distribution $ p_{\rm si}(n_{\rm s},n_{\rm i}) $ of
the weak TWB is then expressed as a two-fold convolution of these three
photon-number distributions \cite{PerinaJr2013a}:
\begin{eqnarray}  
 p_{\rm si}(n_{\rm s},n_{\rm i}) &=& \sum_{n=0}^{{\rm min}(n_{\rm s},n_{\rm i})}
  p_{\rm s}(n_{\rm s}-n;m_{\rm s},b_{\rm s}) \nonumber \\
  & & \times p_{\rm i}(n_{\rm i}-n;m_{\rm i},b_{\rm i})
  p_{\rm p}(n;m_{\rm p},b_{\rm p});
\label{D2}
\end{eqnarray}
$ n_{\rm s} $ ($ n_{\rm i} $) gives the number of signal (idler)
photons. Considering the experimental TWBs whose photocounts form
the signal and idler channels discussed in Appendix A, we revealed
the following parameters \cite{PerinaJr.2021}: $ m_{\rm p} =
m_{\rm s} = m_{\rm i} = 10 $, $ b_{\rm p} = 9.8 \times 10^{-3}$,
$ b_{\rm s} = 3.6 \times 10^{-4}$ and $ b_{\rm i} = 3.9 \times
10^{-5}$.

Photon-number distribution $ p_{\rm si}^{(w_{\rm p})} $ of a TWB
composed of $ w_{\rm p} $ weak TWBs from one double detection window can be
expressed in the form of Eq.~(\ref{D2}) when we suitably
substitute the numbers of modes $ m_{\rm p,s,i} \rightarrow w_{\rm
p} m_{\rm p,s,i} $:
\begin{eqnarray}  
 p_{\rm si}^{(w_{\rm p})}(n_{\rm s},n_{\rm i}) &=& \sum_{n=0}^{{\rm min}(n_{\rm s},n_{\rm i})}
  p_{\rm s}(n_{\rm s}-n;w_{\rm p}m_{\rm s},b_{\rm s}) \nonumber \\
  & & \hspace{-10mm} \times p_{\rm i}(n_{\rm i}-n;w_{\rm p}m_{\rm i},b_{\rm i})
  p_{\rm p}(n;w_{\rm p}m_{\rm p},b_{\rm p}).
\label{D3}
\end{eqnarray}
Symmetrization of photon-number distribution $ p_{\rm si}^{(w_{\rm
p})} $ of this TWB results in the photon-number distribution $
p_{\rm sym}^{(w_{\rm p})} $:
\begin{eqnarray}  
 p_{\rm sym}^{(w_{\rm p})}(n_{\rm s},n_{\rm i}) &=& \sum_{l_{\rm s}=0}^{n_{\rm s}}
   \sum_{l_{\rm i}=0}^{n_{\rm i}}  p_{\rm si}^{(w_{\rm p})}(n_{\rm s}-l_{\rm s},n_{\rm i}-l_{\rm i})
     p_{\rm si}^{(w_{\rm p})}(l_{\rm i},l_{\rm s}). \nonumber \\
  & &
\label{D4}
\end{eqnarray}

The photon-number distribution $ p_{\rm corr}^{(w_{\rm p})} $ of
the correlated part of the analyzed fields is then built from the
symmetric photon-number distributions $ p_{\rm sym}^{(w_{\rm p})}
$ along the formula:
\begin{eqnarray}  
 p_{\rm corr}^{(w_{\rm p})}(n_1,n_2,n_3) &=& \sum_{l_1=0}^{n_1}
  \sum_{l_2=0}^{n_2} \sum_{l_3=0}^{n_3} p_{\rm sym}^{(w_{\rm p})}(n_1-l_1,l_2) \nonumber \\
  & & \hspace{-20mm} \times
  p_{\rm sym}^{(w_{\rm p})}(n_2-l_2,l_3)
  p_{\rm sym}^{(w_{\rm p})}(n_3-l_3,l_1).
\label{D5}
\end{eqnarray}

Photon-number distribution $ p_{\rm n}^{(w_{\rm n})} $ of the noise
part in one beam built from 2 groups of $ w_{\rm n}/2 $ windows (even $ w_{\rm n}$ is assumed)  
is obtained as a convolution of two marginal distributions, one originating
in the signal windows, the other in the idler windows:
\begin{eqnarray}  
 p_{\rm n}^{(w_{\rm n})}(n) &=&  \sum_{n_{\rm s}=0}^{n}
  p_{\rm n, s}^{(w_{\rm n}/2)}(n_{\rm s}) p_{\rm n, i}^{(w_{\rm n}/2)}(n-n_{\rm s}), \\
 p_{\rm n, s}^{(w_{\rm n}/2)}(n_{\rm s}) &=&  \sum_{n_{\rm i}=0}^{\infty}
  p_{\rm si}^{(w_{\rm n}/2)}(n_{\rm s},n_{\rm i}), \nonumber \\
 p_{\rm n, i}^{(w_{\rm n}/2)}(n_{\rm i}) &=&  \sum_{n_{\rm s}=0}^{\infty}
  p_{\rm si}^{(w_{\rm n}/2)}(n_{\rm s},n_{\rm i})  . \nonumber
\label{D6}
\end{eqnarray}

Finally, the photon-number distribution $ p^{(w_{\rm p},w_{\rm
n})}(n_1,n_2,n_3) $ of a 3-beam symmetric Gaussian state composed
of $ w_{\rm p} $ correlated units and $ w_{\rm n} $ noise units
is attained by convolving the correlated- and noise-parts of the
photon-number distributions:
\begin{eqnarray}  
 p^{(w_{\rm p},w_{\rm n})}(n_1,n_2,n_3) &=& \sum_{l_1=0}^{n_1}
  \sum_{l_2=0}^{n_2} \sum_{l_3=0}^{n_3} p_{\rm n}^{(w_{\rm n})}(n_1-l_1) \nonumber \\
 & & \hspace{-30mm} \times  p_{\rm n}^{(w_{\rm n})}(n_2-l_2)
  p_{\rm n}^{(w_{\rm n})}(n_3-l_3) p^{(w_{\rm p})}_{\rm corr}(l_1,l_2,l_3) .
\label{D7}
\end{eqnarray}

\end{document}